\documentclass{aastex}
\usepackage[twocolumn]{emulateapj5}
\usepackage{epsfig}
\usepackage{apjfonts}
\submitted{Accepted for publication in ApJ}

\shorttitle{Circular Polarization of 40 Blazars}
\shortauthors{Homan, Attridge, \& Wardle}

\begin{document}

\title{Parsec-Scale Circular Polarization Observations of 40 Blazars}

\author{Daniel C. Homan\altaffilmark{1}, Joanne M. Attridge\altaffilmark{2}, 
and John F. C. Wardle\altaffilmark{1}}
\altaffiltext{1}{Physics Department MS057, 
                 Brandeis University, Waltham, MA 02454; 
                 e-mail: dhoman@brandeis.edu,
                         jfcw@quasar.astro.brandeis.edu}
\altaffiltext{2}{MIT Haystack Observatory,
                 Westford, MA 01886; 
                 e-mail:jattridge@haystack.mit.edu}

\begin{abstract}
We present circular polarization results from a 5 GHz survey 
of the parsec-scale polarization properties of 40 AGN made with the VLBA.  
We find 11 circular polarization detections at the $3\sigma$ level or 
higher. This nearly quadruples the number
of sources detected in circular polarization at VLBI resolution.
We find no correlation between fractional linear 
and circular polarization across our sample. A likely explanation
is external Faraday depolarization in the cores of AGN which 
reduces linear polarization but leaves circular polarization 
unchanged.
In comparing ours and other recent results to observations
made $\sim 20$ years ago, we find that, in five of six cases, 
sources have the same sign of circular polarization today as they 
did $20$ years ago. This suggests the presence of a long
term property of the jets, such as the polarity of a net 
magnetic flux, which is stable on time-scales much longer 
than those of individual outbursts.  
\end{abstract}


\section{Introduction}
\label{s:intro}

Circular polarization (CP) observations of radio-loud active 
galactic nuclei (AGN) provide a unique opportunity, especially
in conjunction with linear polarization observations, for 
constraining the magnetic field geometry, particle energy
distribution, and even particle composition ($e^+e^-$ vs. $p^+e^-$) 
of their jets.  A number of possible mechanisms 
have been proposed for the production of circular polarization 
in AGN (see \citet{WH01} and references therein). The two most 
likely are the {\em intrinsic} circular polarization of
synchrotron radiation and Faraday {\em
conversion} of linear to circular polarization, a propagation
effect, e.g. \citet{JOD77}.  \citet{WDP83} 
catalog a large number of circular polarization observations
of AGN, mainly from the 1970s, and find typical levels of 
$\lesssim 0.1$\%.  After the survey by \citet{K84}, 
there have been no major CP observations published until 
recently. 

In \citet{WHOR98}, we reported the detection of circular 
polarization in the parsec-scale jet of 3C\,279.  Detailed 
analysis suggested 
the jet is predominately a pair plasma.  Circular polarization 
has since been detected in the galactic center \citep{BFB99,SM99},
the galactic X-ray binary SS433 \citep{F00}, the intra-day
variable source PKS 1519-273 \citep{M00}, and a sample
of more than twenty AGN by \citet{RNS00}, hereafter RN\&S.
In \citet{HW99}, hereafter H\&W, we presented multi-epoch,
parsec-scale CP detections of four AGN, made with the
Very Long Baseline Array (VLBA)\footnote{The National Radio Astronomy
Observatory is a facility of the National Science Foundation
operated under cooperative agreement by Associated Universities, Inc.}
at 15 GHz.  

Here we present results from the first large scale circular
polarization survey made at Very Long Baseline Interferometry 
(VLBI) resolution. In December 1996, we observed a sample of 40 AGN at 
5 GHz ($\lambda$ 6 cm) with the VLBA.  These 
observations were originally made as part of a continuing 
study of the parsec-scale linear polarization properties of 
blazars (\citet{A99}; Attridge et al., in prep.). They do 
not comprise a complete sample.  We report the results of a 
re-analysis of these data for circular polarization.

%
%
%

\section{Observations and Calibration}
\label{s:obs}

The observations were scheduled in a continuous 48 hour block, 
with source changes after every 5.5 minute scan.  Sources were 
highly interleaved in the schedule to maximize (u,v)-coverage. 
Each source was observed for at least ten scans, five on each day, 
giving a total integration time of nearly an hour\footnote{The feed
leakage term calibrator, OQ\,208, was observed for ten scans on each
day to assure outstanding parallactic angle coverage.}.
The data were recorded at each station using 1-bit sampling and 
were correlated at the VLBA correlator in Socorro, NM.  
The correlated data contained all four cross-correlations 
(RR, RL, LR, LL), each with four intermediate frequencies (IFs).  
The data were loaded into NRAO's Astronomical Imaging
Processing System (AIPS) \citep{BG94,G88} and calibrated using standard
techniques for VLBI polarization observations, e.g., \citep{C93, RWB94}.
The VLBA station in Hancock, NH was found to have a low flux 
calibration and significantly higher noise than the other antennas 
and was dropped from the observations.  For a specific description 
of the original calibration steps see \citet{A99}.

The reader is referred to H\&W for a detailed discussion of how
to detect small levels of circular polarization with the VLBA.
There we describe three techniques for robustly detecting 
circular polarization with circularly polarized antenna feeds.  
In our re-analysis of the 5 GHz survey for circular polarization, 
we applied the {\em gain transfer} technique which provides a 
direct measure of the circular polarization in a source.   
We were unable to apply the other techniques described in H\&W due to 
both a lack of strong circular polarization signals in the sample 
(most are a few mJy or less) and the lack of sufficiently bright 
extended structure. 

The application of the gain transfer technique is straight forward, 
requiring that we simply make no assumption about the circular polarization 
of a source during self-calibration, i.e. we only assume 
$(RR+LL)/2 = \tilde{I}_{model}$.
This will leave uncalibrated the $R/L$ complex gain ratio at each 
antenna.  For the source model, we used the best model obtained from the
original analysis\footnote{In a few cases, we found that better source 
models could be made, and in those cases, we used the revised models.}, 
and self-calibrated the data in amplitude and phase.  After self-calibration,
we removed the effects of antenna feed leakage determined by applying
the AIPS task PCAL to the unpolarized source OQ208.  This step is crucial 
as uncorrected feed leakage terms may induce non-closing errors in the
parallel hand data.  After feed leakage correction, we self-calibrated 
the data again in amplitude and phase using the same model as before.  
The data were then carefully 
edited to remove clearly discrepant polarization data from stokes Q, U, 
and V in an antenna based manner.\footnote{The careful editing in total 
intensity done during the original analysis was preserved.  Edits
were always applied to all four cross-correlations.}  These ``final'' 
data, could then be used to either (1) solve for the $R/L$ complex 
gain ratios by self-calibration assuming no 
circular polarization in that source ($RR=\tilde{I}_{model}$, 
$LL=\tilde{I}_{model}$) , or (2) look for circular polarization 
on a source by applying a smoothed set of $R/L$ complex gains from a 
sub-set of sources.

The key to gain transfer calibration is determining a smoothed set of 
$R/L$ complex antenna gains. As suggested in H\&W, we initially used 
nearly the entire 40 source sample to produce a smoothed set of gains.  
The large sample size was a great asset, and by smoothing the 
gains over several hours and averaging over many sources, we 
were insensitive to the particular characteristics (such
as strong circular polarization) of any one source.  We then applied this
smoothed set of gains back to the sources, imaged in circular polarization,
and removed the most strongly polarized sources from contributing to the
smoothed gains.  We repeated this process through a few iterations 
until there were no sources with apparent CP $\geq 0.12$\% that
contributed to the smoothed gains.\footnote{This includes eliminating
14 weak sources ($< 800$ mJy) which would not produce at least
$1$ mJy of CP at $0.12$\%.}  This cutoff gave thirteen sources 
contributing to the smoothed gains with a boxcar averaging window 
of six hours.  Our main results are insensitive to both the details 
of which sources contribute to the gains, and our choice of smoothing 
function and time-scale. 

The reader is referred to H\&W for a detailed discussion of the
uncertainties associated with circular polarization observations with
circularly polarized feeds.  Here we would like to briefly remind the
reader of two points. (1) While it is important to remove 
feed leakage terms from the parallel hand data, contributions 
from the small uncertainties in the determination of 
the feed leakage terms are insignificant relative to the gain 
uncertainties. (2) Beam squint, resulting from pointing
errors, shows up as a pure amplitude gain effect and is therefore 
naturally included in our estimate of gain uncertainties.  In the
appendix we have refined our estimate of the gain calibration 
uncertainties as presented in H\&W. 

As a final check on our results, we divided the single 48-hour data 
set into two sub-sets and imaged each separately in CP.  This provided
a direct test of our understanding of the uncertainties from
all but the longest time-scale effects ($> 12$h).  
We found no significant differences 
between the two parts of the 48-hour observation.  

\section{Results}
\label{s:results}

\begin{figure}
\figurenum{1}
\epsscale{0.46}
\plotone{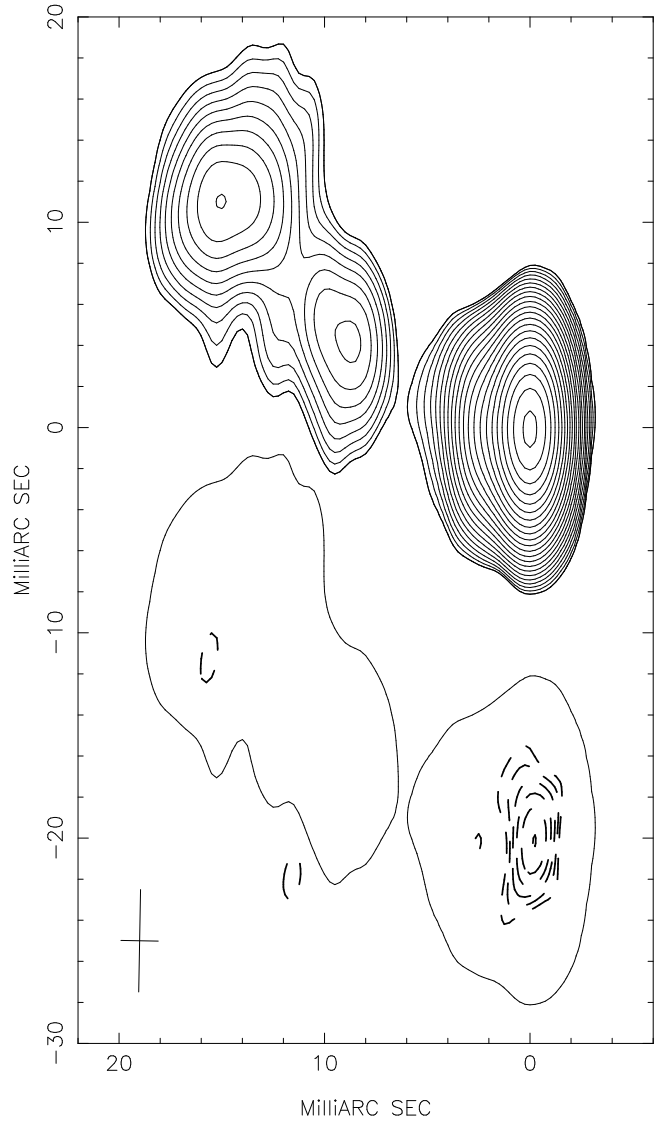}
\caption[f1.eps]{\label{f:0607}Total intensity (lighter
contours; factor of $\sqrt{2}$ beginning at $2$ mJy/beam) 
and circular polarization (darker contours; factor of 
$\sqrt{2}$ beginning at $\pm1$ mJy/beam; dashed contours 
are negative) of the parsec scale jet of the quasar 
PKS 0607-157 at 5 GHz, epoch 1996.96.  A single Stokes $I$ contour
is overlayed with Stokes $V$ to show the registration.
The peak circularly polarized flux is $V = -5.7\pm1.6$ mJy/beam, 
$V/I = -0.18\pm0.05$\%.}
\end{figure}

Table \ref{t:results} lists our results,
and Figure \ref{f:0607} presents an image of the circular
polarization distribution of a typical source (PKS 0607-157).  
The sources in this sample tend to be strongly
core dominated, and the CP we observe is coincident 
with the bright VLBI cores. Any small displacements from
the VLBI cores do not appear significant within our 
positional uncertainties which are approximately a 
beam-width divided by the SNR of the CP measurement.  
For this reason, we simply report 
the measured CP as a fraction of the peak total intensity.  
The uncertainties and limits listed 
in Table \ref{t:results} are dominated by gain calibration
uncertainties which are estimated in the appendix.
The total quoted uncertainty, $\sigma$, includes both 
the gain uncertainty and the RMS noise in the stokes V map,
added in quadrature. The limits in Table \ref{t:results} 
are the largest of $1$~mJy, $|V_{peak}|+1\sigma$, or
$2\sigma$. 



\section{Discussion}
\label{s:discuss}

\subsection{Detection Rate of Circular Polarization}
\begin{figure}
\figurenum{2}
\epsscale{0.7}
\plotone{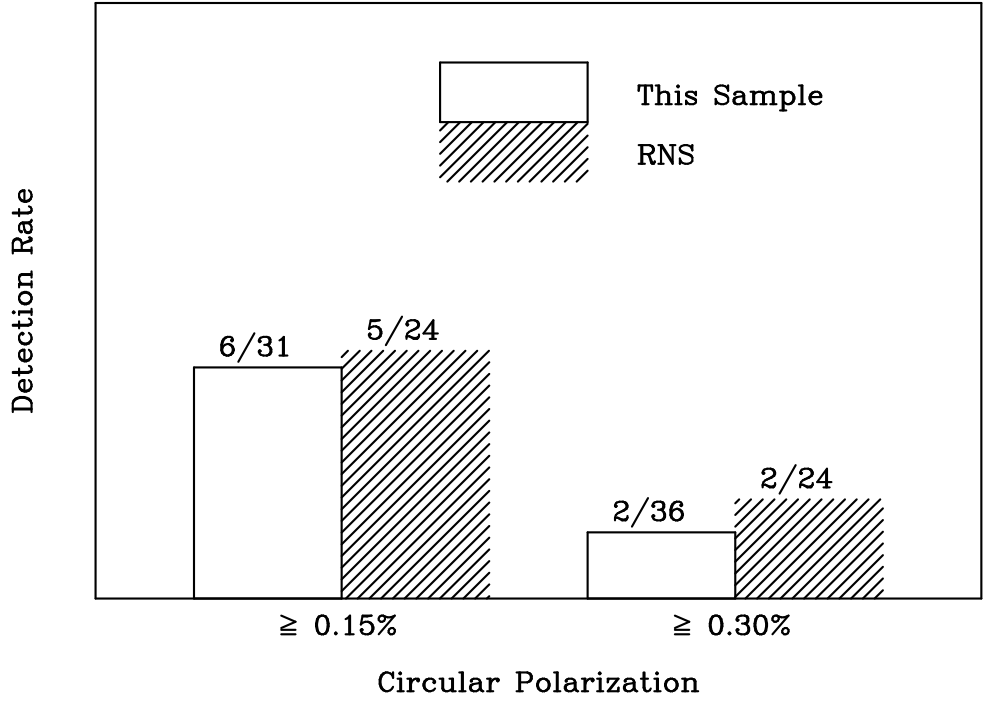}
\caption[f2.eps]{\label{f:detect}Detection rates of circular 
polarization among
BL Lacs and quasars.  Sources in our sample which are not strong 
enough to produce a CP signal of at least 1 mJy/beam at a given 
fractional level are excluded.}
\end{figure}

Figure \ref{f:detect} compares our detection rates for
circular polarization for BL Lacs and quasars with the
sample of RN\&S (also at 5 GHz).  For this comparison we
have divided by
the total VLBI flux (rather than the peak VLBI flux) to better 
compare to the Australian Telescope Compact Array (ATCA) 
integrated measurements.  Our detection 
rates are very similar to those of RN\&S.  


We can do a similar comparison to the 15 GHz VLBI sample
of H\&W.  Here we compare the peak values for fractional
circular polarization, $m_c$, in our 5 GHz measurements 
to $m_c$ of the integrated core at 15 GHz, including model 
components out to 1.5 mas.  
This comparison essentially eliminates resolution based 
differences between the samples.  While we find only 2/36 
quasars or BL Lacs with $|m_c| \geq 0.3$\% in the 5 GHz 
measurements presented here, the 15 GHz sample of H\&W has
a significantly higher detection rate for strong CP.  
There are five closely spaced ($\sim$ 2 month intervals) epochs
in H\&W which include the same 11 quasars and BL Lacs, three
of which show repeated detections of CP.
Averaging over the epochs in H\&W, we have a mean of 2.4/9.6 
detections of $|m_c| \geq 0.3$\% from the quasars and BL Lacs 
strong enough in total intensity in those epochs to have 
detectable CP at this level.  Although the total number
of sources in H\&W is small, this detection rate is nearly 
five times that seen here at 5 GHz, and we note that most of the 
detections in H\&W are $\geq 0.5$\% CP, and we observe 
no source with CP that strong in this 5 GHz sample.

The higher incidence of strong circular polarization 
detections in H\&W may be due to one of three possible 
causes: (1) random 
fluctuations in small samples, (2) a broad frequency 
dependent effect, or (3) source selection criteria.
We can evaluate option (1) by assuming simple binomial 
statistics based on our 5 GHz observations (i.e. assume
the chance of observing $|m_c| \geq 0.3$\% is $\approx (2\pm\sqrt{2})/36$).  
We find we would need approximately a 2.5 sigma random 
variation to explain the 15 GHz results of H\&W.  
Any broad frequency dependent 
effects, option (2), depend on the generation 
mechanism of the circular polarization.  While both intrinsic
CP and Faraday conversion have specific, frequency dependent
signatures, the inhomogeneous, self-similar nature of VLBI
radio cores complicates the analysis considerably (e.g. 
\citet{J88}) and may lead to a flat $m_c$ spectrum for
both mechanisms \citep{WH01}.

With no clear frequency dependent effect, we suspect that 
source selection criteria, option (3) above,
may explain the higher levels of circular polarization 
observed in H\&W. 
The sample of H\&W was part of multi-epoch 
monitoring program of the strongest, most currently 
active blazars (\citet{H01}; Ojha et al. in prep.), and the
three quasars with high levels of CP were undergoing violent
core outbursts during the observations.  This is consistent 
with the report by RN\&S of a significant correlation
between overall source variability and fractional CP.  For RN\&S, who 
have only integrated observations, the correlation with 
source variability may be confused by a correlation with overall
source ``compactness'', as extended structure will 
decrease both the fractional CP of a source and its fractional
variability.  For the VLBI observations of H\&W compactness is
not an issue, and the higher levels of fractional CP appear
correlated with the selection criteria of currently active 
sources for that sample. 

It is unclear how current core outburst events may lead to
increased fractional circular polarization via the intrinsic
mechanism.  Outbursts are typically associated with shocks
that tend to order the magnetic field and increase particle
densities. Increased particle densities should not
enhance the {\em fractional} contribution of intrinsic circular 
polarization, and it is difficult to imagine how the 
{\em uni-directional} field necessary for intrinsic CP should be 
preferentially enhanced by a shock. If the uni-directional
field is along the jet axis and the shocks are transverse, 
just the opposite should happen, and the relative contribution
from intrinsic circular polarization should decrease. 
For Faraday conversion, ordered magnetic fields are necessary 
for both the generation of linear polarization and its conversion 
to circular polarization.  Shocks in the core may very well 
increase the contribution of Faraday conversion to the circular 
polarization produced. 

\subsubsection{Distribution with Source Type}

RN\&S found no significant difference between the distribution of
circular polarization in the quasars and BL Lacs in their sample, although
they did find that BL Lacs and quasars had significantly higher levels
of CP than radio galaxies.  Their detailed
analysis was made possible by the great precision ($\pm0.01$\%)
with which ATCA can measure integrated CP.  Due to our higher 
uncertainty ($\sim\pm0.05$\%), which 
induces a dependence on source strength, we cannot do a similar 
detailed analysis.  The BL Lacs in our sample tend
to be weaker than the quasars and many have upper limits higher than
the typical detection level for quasars.

\subsection{Correlation with Linear Polarization}

\begin{figure}
\figurenum{3}
\epsscale{0.7}
\plotone{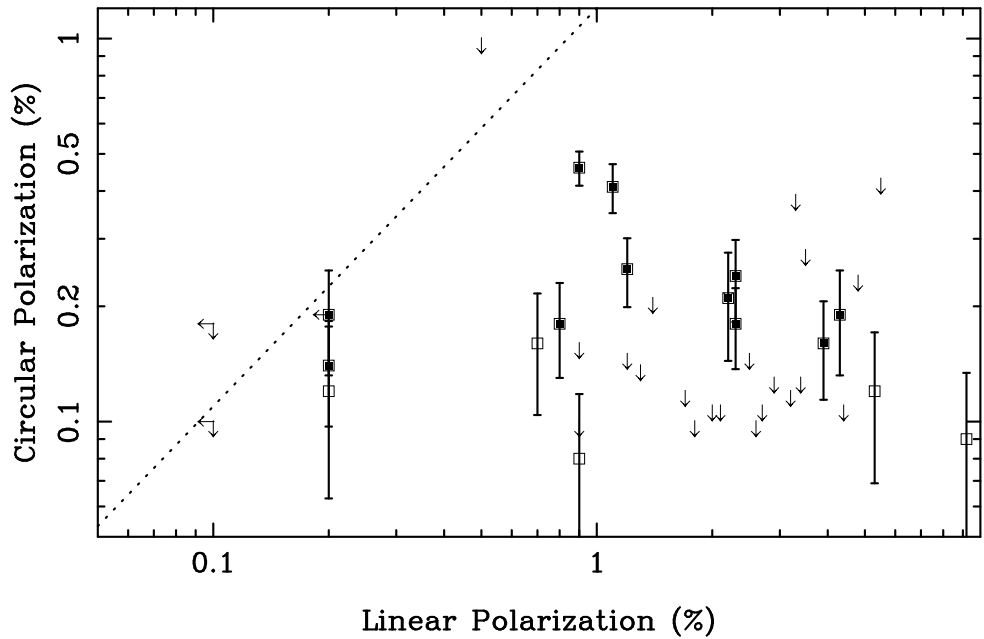}
\caption[f3.eps]{\label{f:mc_ml}Plot of (absolute) fractional core circular 
polarization versus fractional core linear polarization.  The dashed line 
indicates $|m_c| = m_l$.  Filled squares indicate $\geq 3\sigma$ detections.  
Upper limits are represented by arrows.}
\end{figure}

Figure \ref{f:mc_ml} is a plot of fractional core circular 
polarization, $|m_c|$, versus fractional core linear polarization, 
$m_l$.  Beyond the fact that linear polarization is almost 
always stronger than circular polarization, there does not appear to 
be any correlation between linear and circular polarization of
the core.  This was also noted by RN\&S.  One might expect that 
fractional linear polarization, a measure of field order, should 
correlate with fractional CP, as disordered magnetic fields will not 
generate significant {\em net} CP by either the intrinsic or 
conversion mechanism.
The issue is complicated, however, by the possibility of 
Faraday depolarization in the cores of these objects.  Large rotation 
measures have been observed in the cores of quasars by
\citet{T98,T00}, and if the Faraday effect is occurring in an external
screen, the CP will remain largely unmodified\footnote{Thermal particles
are inefficient at Faraday conversion compared to Faraday 
rotation, $\tau_{rot}/\tau_{con} \sim 2\times10^5(\lambda B)^{-1}$ (cgs), 
e.g. \citet{JOD77}}, 
although the linear polarization may be significantly reduced.  
The scintillation mechanism described by \citet{MM00} for the
production of CP has no correlation with linear polarization, 
but is expected to have rapid changes ($\sim$ minutes--hours) 
in sign which are not observed in AGN (see \S{\ref{s:consist}}).


We find no correlation for fractional CP with either total 
intensity or redshift in this sample.

\subsection{Comparison to Other Observations}

\subsubsection{Recent Observations}
We have a single source in common with the observations 
of RN\&S: PKS 0454-234.  They found 
$+0.26\pm0.01$\% integrated circular polarization in 1997.17 at 5 GHz, 
compared to our $2.8\sigma$ result of $+0.16\pm0.06$\% core circular 
polarization measured 11 weeks earlier (also at 5 GHz).  
For the source PKS 0607-157 we detect circular polarization of 
approximately $-0.7$\% in January of 1998 at 8 GHz 
(Homan et al., in prep.), a signal which is stronger, but
of the same sign as we report here at 5 GHz in observations made
a full year earlier.  This kind of short-term sign consistency in
CP observations, as well as the possibility
for longer-term consistency, is discussed in 
\S{\ref{s:consist}}.

In 3C\,279 we observe strong circular polarization
for five epochs in 1996 at 15 GHz (H\&W); however, in the 5 GHz sample
presented here, we do not detect significant CP for 3C\,279 
in December of 1996 -- a little more than two months after the 
final 15 GHz CP observations by H\&W.  In \citet{WHOR98} we argue
that the CP in 3C\,279 is associated with
the core-west (CW) component which appears to be part of a 
core outburst event in 3C\,279 during 1996.  This component would
have been strongly self-absorbed at 5 GHz in late 1996, so we 
should not expect to see any CP from it. 

\subsubsection{Long-term Sign Consistency}
\label{s:consist}
\citet{K84} noted that circularly polarized sources tended to
have a {\em preferred handedness} or {\em sign} which they maintained
over the course of a few years at 5 GHz.  In H\&W, we observed the
same effect at 15 GHz over five epochs, taken at two month intervals 
during 1996.  Here we present evidence that this sign consistency
may persist for decades.  Between our sample and those of 
RN\&S and H\&W, there are six sources with circular polarization detected 
both recently and $\sim 20$ years ago \citep{K84,HA77,WDP83}: 3C\,84, 
PKS 0537-441, 3C\,273, 3C\,279, PKS 1921-293, and PKS 1934-638.  
Of those six sources, five have the same sign of circular polarization 
today as they preferred two decades ago\footnote{It is interesting to 
note that two of our $2\sigma$ measurements, CTA 102 and 3C\,454.3,
were also detected $\sim 20$ years ago with the same sign \citep{WDP83}.}.

Though the statistics are small, this result is potentially very 
interesting.  It suggests a long-lived property of the jets, stable on 
the time-scale of multiple outbursts, which sets the preferred sign of 
circular polarization \citep{HW99,WH01}.  A natural possibility for
this property is the polarity of a net magnetic flux in the jet, 
reflecting the magnetic flux at the central engine. 
This is a fundamental parameter in electro-magnetic models of 
jet production (e.g. \citet{BP82, LR95}).  A net flux can
generate either intrinsic circular polarization or
may drive the Faraday conversion process with a small amount
of Faraday rotation, e.g. \citet{WHOR98}.  
Another possible long-term property is the handedness of a helical 
field which may produce significant quantities of 
Faraday conversion, e.g. \citet{H82}.

\section{Conclusions}
\label{s:conclude}
We have presented circular polarization results from a 
survey of the parsec-scale polarization polarization of 
40 AGN at 5 GHz.  We found 11 circular polarization
detections at the $3\sigma$ level or higher.  This 
nearly quadruples the number of sources detected
in CP at VLBI resolutions.  We observed only two sources with high 
levels of CP, $|m_c| \geq 0.30$\%, and we believe that 
current outbursts in the sources observed by H\&W could be related
to the higher levels of CP detected there.  We found no 
correlation between fractional linear 
and circular polarization across our sample.  This may be due
to Faraday depolarization in the cores of AGN which will leave
circular polarization unaffected. In comparing these and other 
recent results to CP observations made $\sim 20$ years ago, 
we found that, in five of six cases, sources have the same 
sign of CP today as they did $20$ years ago. This suggests 
the presence 
of a long term property of the jets, such as the polarity of
a net magnetic flux, that is stable on the time-scale of 
multiple outbursts.  

\section{Acknowledgments}

This work has been supported by NSF Grants AST 95-29228 and 
AST 98-02708.  This research has made use of the NASA/IPAC 
Extragalactic Database (NED) which is operated by the Jet 
Propulsion Laboratory, California Institute of Technology, 
under contract with the National Aeronautics and Space 
Administration. This research has also made use of NASA's 
Astrophysics Data System Abstract Service.


\appendix

\section{Uncertainties from Gain Transfer}
\begin{figure}
\figurenum{4}
\plotone{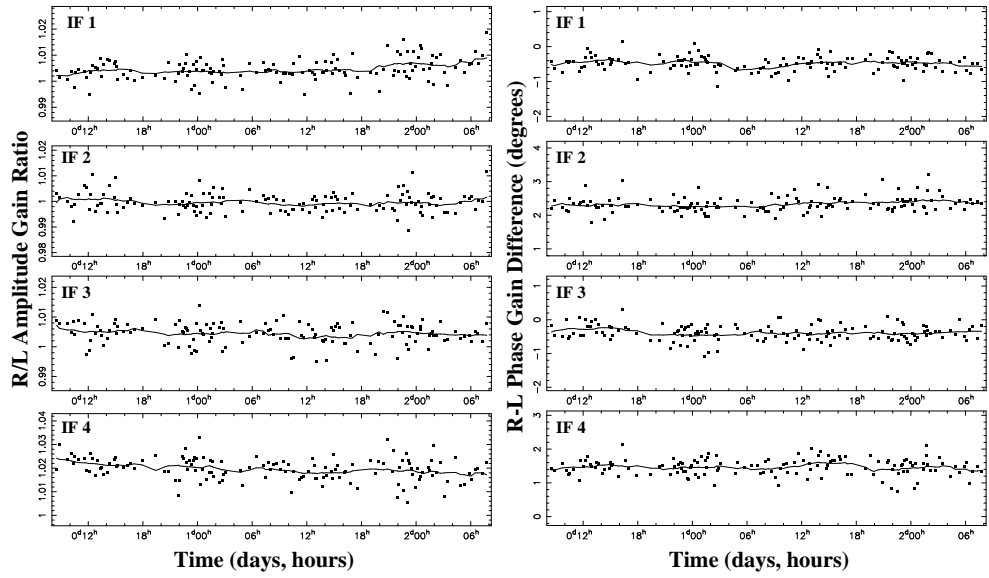}
\caption[f4.eps]{\label{f:BRgains}R/L amplitude 
gain ratio corrections (left) and
R-L phase gain corrections (right) at the Brewster, WA VLBA station.  These 
are the corrections for the thirteen calibrator sources, assuming 
zero circular polarization during the final self-calibration (e.g. 
$RR = \tilde{I}_{model}$, $LL = \tilde{I}_{model}$).  A six-hour sliding
boxcar average is plotted as a solid line through the points.}
\end{figure}

The calibration of circular antenna feeds for circular
polarization is described in detail in H\&W. 
The {\em gain transfer} procedure allows calibration of the $R/L$ 
complex antenna gain ratio\footnote{The average gain corrections
at each antenna have already been removed by earlier rounds of 
self-calibration which make no assumption about the presence
or absence of CP, i.e. they only assume $(RR+LL)/2 = \tilde{I}_{model}$.} 
at each antenna by self-calibration assuming zero circular 
polarization ($\tilde{RR} = \tilde{I}_{model}$, 
$\tilde{LL} = \tilde{I}_{model}$) on a subset of sources, smoothing the
derived antenna gains, and applying the smoothed gains to all sources.
Figure \ref{f:BRgains} shows these gains for a typical antenna in our
observations.


In H\&W, we showed that there were two main contributions
to the complex antenna gain ratios: (1) a long-term offset and (2) a
short-term rapid variation.  Both contributions were of the order of 
a percent, and while the long-term offset could be easily corrected, 
the short-term rapid variation went essentially uncorrected.  We noted
that the short-term rapid variations seemed uncorrelated between antenna,
IF, and scan, and used this to crudely estimate our uncertainties from 
gain transfer calibration.
 
In this appendix we provide a more detailed analysis of the uncertainties 
associated with this calibration technique.  For a given antenna, assume we 
have corrected the complex antenna gain ratio, leaving some residual
error $\delta$ in amplitude and $\phi$ in phase.  Under these
circumstances, the residual, uncorrected gain ratio can be written 
for antenna $a$ as
\begin{equation}
\left( \frac{G_R}{G_L} \right)_a = (1+\delta_a)e^{i\phi_a}
\end{equation}
\noindent where both $\delta$ and $\phi$ are on the order of 0.005 with 
the largest contributions from rapid, uncorrelated variations. In the
Stokes $V$ visibilities, these errors show up, to first order, multiplying
the total intensity:
\begin{equation}
\tilde{V}_{ab} = \tilde{V}_{true} + \tilde{I}_{ab}[(\delta_a/2+\delta_b/2)
+i(\phi_a/2-\phi_b/2)]
\end{equation}

Here we consider amplitude  
errors, as only they show up on the phase center and may be confused for
a real circularly polarized signal. For point-like sources, phase errors 
will not contribute to first order, and for sources with significant 
extended structure, phase errors will appear in an anti-symmetric 
fashion about the phase center.  

The total uncertainty from gain transfer, $\sigma_{gains}$, in the 
fractional circular polarization includes contributions to $\delta$ from 
three sources: 
(1) uncertainty in the determination of the smoothed antenna gains, 
(2) uncertainty in the true circular polarization of the calibrator 
sources, and (3) uncorrected rapid variations in the antenna gain 
ratios.

\paragraph{Uncertainty in the smoothed antenna gains} is estimated
by examining the scan to scan variations of the thirteen sources used
to determine the average gains.  We computed the RMS variation in the 
individual gains from the six hour boxcar average determined on these 
sources.  Before computing the RMS, we took a direct average of 
the variations across all IFs for each scan.  This is to avoid having
to make any assumptions about the degree of correlation in the variations 
between IFs\footnote{This RMS result was roughly $\sqrt{2}$ higher than 
simply assuming the four IFs were completely uncorrelated in their 
variations.}.
We found an RMS variation for the calibrators of 
$(\delta_{cal})_{RMS} = 0.0029$.  The uncertainty in any given six hour 
average is given by $(\delta_{cal})_{RMS}/\sqrt{N_{scans}}$, where $N_{scans}$
is the number of scans in that averaging period, and the 
contribution to the circular polarization of a target source is
\begin{equation}
\sigma_{avg} = \frac{(\delta_{cal})_{RMS}}{\sqrt{N_{scans}N_{ant}N_{avg}}}.
\end{equation}
\noindent For these observations the number of antennas, $N_{ant}$,
was nine.  The number of independent averages, $N_{avg}$, which apply
to a given source was taken to be two as each source 
was observed over a period of five to six hours on each of the 
two days of the experiment. The average numbers of 
calibrator scans, $N_{scans}$, contributing to the gains applied
to each source were computed and are listed in Table \ref{t:errors}.

\paragraph{Uncertainty due to real CP on the calibrators} is 
easily estimated by finding the RMS apparent CP of the calibrators
and dividing by the square root of the number of calibrators 
contributing to a given gain average:
\begin{equation}
\sigma_{res} = (m_c)_{RMS}/\sqrt{N_{cal}}.
\end{equation}
\noindent For these observations, the RMS peak circular polarization
of the calibrators was $(m_c)_{RMS} = 0.00065$.  As each source (targets
and calibrators alike) was observed over a five to six hour period on
each day, and the gain averages are done over a six hour interval, each 
source overlaps with many calibrator sources.  Unfortunately,
the contributions of each of these calibrator sources were not equal, and we
needed to use a weighted average computed from the experiment schedule
and listed in Table \ref{t:errors}.  These weighted averages were
determined in a conservative fashion: the calibrator source with the largest 
contribution (number of calibrator scans that influenced each target scan) 
to a given source was treated as a 1.0 contribution, and the other calibrators
had their {\em relative} contribution added to this.

\paragraph{Uncertainty due to random scan to scan gain variations} can be
directly computed for each source:
\begin{equation}
\sigma_{ran} = \frac{(\delta_{source})_{RMS}}{\sqrt{N_{scans}N_{ant}}}.
\end{equation}
\noindent The RMS variation for each source was computed relative
to a twelve-hour average on that source, averaging the variations across 
IF before computing the RMS so that we made no assumption about
the extent of correlated variation between IF\footnote{For sources with
large scan to scan variations, we obtained essentially the same result
if we simply assumed the four IFs were completely uncorrelated, but for
sources with smaller variations, the result was approximately 
$\sqrt{2}$ smaller if we assumed the IFs were uncorrelated in their
variations.}.


Table \ref{t:errors} lists our results for each of these contributions
on every source as well as the total uncertainty, $\sigma_{gains} = 
\sqrt{\sigma^2_{avg}+\sigma^2_{res}+\sigma^2_{ran}}$, 
contributed by our gain transfer calibration.



\newpage


\begin{deluxetable}{ccccccccc}
\tablewidth{0pc}
\tablecolumns{9}
\tabletypesize{\scriptsize}
\tablenum{1}
\tablecaption{Circular Polarization Results.\label{t:results}}
\tablehead{$Object$ & $Alt.$ $Name$ & $Redshift$ & $Class$ & $I_{peak}$ & $m_l$ & $V_{peak}$ 
& $m_c$ & $\sigma$ \\
(B1950) &&&& (mJy/beam) & (\%) & (mJy/beam) & (\%) & }
\startdata
0215$+$015 & PKS & $1.715$ & BL &  $477$ & $1.4$ & \nodata & $< 0.21$ & \nodata \\
0219$-$164 & PKS & $0.698$ & BL &  $436$ & $2.2$ & \nodata & $< 0.23$ & \nodata \\
0219$+$428 & 3C\,066A & $0.444$ & BL &  $792$ & $2.9$ & \nodata & $< 0.13$ & \nodata \\
{\bf 0336$-$019} & {\bf CTA 26} & {\boldmath $0.852$} & {\bf Q} & {\boldmath $1069$} & {\boldmath $4.3$} 
                 & {\boldmath $+2.02\pm0.62$} & {\boldmath $+0.19\pm0.058$} & {\boldmath $3.3$} \\
0403$-$132 & PKS & $0.571$ & Q & $639$ & $0.9$ & \nodata & $< 0.16$ & \nodata \\ 
0420$-$014\tablenotemark{a} & PKS & $0.915$ & Q & $2005$ & $2.7$ & \nodata & $< 0.11$ & \nodata \\
0422$-$004 & PKS & $0.310$ & BL &  $357$ & $3.5$ & $-0.71\pm0.28$\tablenotemark{b} & $< 0.28$ & \nodata \\
0454$-$234 & PKS & $1.009$ & Q & $1247$ & $0.7$ & $+1.96\pm0.70$ & $+0.16\pm0.056$ & $2.8$ \\
0605$-$085\tablenotemark{a} & PKS & $0.870$ & Q & $1832$ & $0.9$ & \nodata & $< 0.10$ & \nodata \\
{\bf 0607$-$157} & {\bf PKS} & {\boldmath $0.324$} & {\bf Q} & {\boldmath $3161$} & {\boldmath $0.8$} 
                 & {\boldmath $-5.72\pm1.57$} & {\boldmath $-0.18\pm0.050$} & {\boldmath $3.6$} \\
{\bf 0743$-$006} & {\bf PKS} & {\boldmath $0.994$} & {\bf Q} & {\boldmath $1489$} & {\boldmath $0.9$} 
                 & {\boldmath $-6.82\pm0.70$} & {\boldmath $-0.46\pm0.047$} & {\boldmath $9.7$} \\
0846$+$513 & & $1.860$ & BL &  $259$ & $3.3$ & \nodata & $< 0.39$ & \nodata \\
0851$+$202\tablenotemark{a} & OJ287 & $0.306$ & BL & $1053$ & $2.6$ & \nodata & $< 0.10$ & \nodata \\
0855$+$143 & 3C\,212 & $1.048$ & Q &  $102$ & $0.5$ & \nodata & $< 1.0$ & \nodata \\
0906$+$015 & DA 263 & $1.018$ & Q &  $234$ & $5.5$ & \nodata & $< 0.43$ & \nodata \\
{\bf 0906$+$430} & {\bf 3C\,216} & {\boldmath $0.670$} & {\bf Q} & {\boldmath $485$} & {\boldmath $< 0.2$} 
                 & {\boldmath $-0.92\pm0.28$} & {\boldmath $-0.19\pm0.058$} & {\boldmath $3.3$} \\
1055$+$018\tablenotemark{a} & PKS & $0.888$ & Q & $1565$ & $3.4$ & $+1.25\pm0.72$ & $< 0.13$ & \nodata \\
{\bf 1150$+$497} & {\bf 4C\,49.22} & {\boldmath $0.334$} & {\bf Q} & {\boldmath $817$} & {\boldmath $1.1$} 
                 & {\boldmath $-3.39\pm0.49$} & {\boldmath $-0.41\pm0.060$} & {\boldmath $6.9$} \\
1156$+$295\tablenotemark{a} & 4C\,29.45 & $0.729$ & Q & $1329$ & $2.1$ & \nodata & $< 0.11$ & \nodata \\
1253$-$055\tablenotemark{a} & 3C\,279 & $0.536$ & Q & $4711$ & $2.0$ & $-2.95\pm2.20$ & $< 0.11$ & \nodata \\
1334$-$127 & PKS & $0.539$ & Q &  $3218$ & $5.3$ & $-3.97\pm1.63$ & $-0.12\pm0.051$ & $2.4$ \\
1404$+$286\tablenotemark{a} & OQ208 & $0.077$ & SyI & $1575$ & $< 0.1$ & \nodata & $< 0.10$ & \nodata \\
1413$+$135 & OQ122 & $0.247$ & BL &  $743$ & $< 0.1$ & $-0.82\pm0.45$ & $< 0.18$ & \nodata \\
1502$+$106 & OR103 & $1.839$ & Q & $1240$ & $3.2$ & \nodata & $< 0.12$ & \nodata \\
1504$-$166 & PKS & $0.876$ & Q &  $1475$ & $2.5$ & $+1.36\pm0.83$ & $< 0.15$ & \nodata \\
1510$-$089 & PKS & $0.360$ & Q & $1043$ & $4.8$ & $-1.64\pm0.82$\tablenotemark{c} & $< 0.24$ & \nodata \\
{\bf 1546$+$027} & {\bf PKS} & {\boldmath $0.413$} & {\bf Q} & {\boldmath $1958$} & {\boldmath $1.2$} 
                 & {\boldmath $-4.88\pm0.99$} & {\boldmath $-0.25\pm0.051$} & {\boldmath $4.9$} \\
1656$+$053 & PKS & $0.879$ & Q &  $796$ & $0.2$ & $+0.95\pm0.45$ & $+0.12\pm0.057$ & $2.1$ \\
{\bf 1842$+$681} & & {\boldmath $0.472$} & {\bf Q} &  {\boldmath $428$} & {\boldmath $2.2$} 
                 & {\boldmath $+0.88\pm0.28$} & {\boldmath $+0.21\pm0.066$} & {\boldmath $3.1$} \\
{\bf 1921$-$293} & {\bf PKS} & {\boldmath $0.352$} & {\bf Q} & {\boldmath $10370$} & {\boldmath $3.9$} 
                 & {\boldmath $-16.3\pm4.8$} & {\boldmath $-0.16\pm0.046$} & {\boldmath $3.4$} \\
1958$-$179\tablenotemark{a} & PKS & $0.650$ & Q & $1736$ & $1.8$ & \nodata & $< 0.10$ & \nodata \\
2032$+$107 & PKS & $0.601$ & BL & $672$ & $1.2$ & \nodata & $< 0.15$ & \nodata \\
{\bf 2201$+$171} & {\bf PKS} & {\boldmath $1.075$} & {\bf Q} & {\boldmath $698$} & {\boldmath $2.3$} 
                 & {\boldmath $+1.66\pm0.41$} & {\boldmath $+0.24\pm0.058$} & {\boldmath $4.0$} \\
{\bf 2223$-$052} & {\bf 3C\,446} & {\boldmath $1.404$} & {\bf Q} & {\boldmath $1373$} & {\boldmath $2.3$} 
                 & {\boldmath $+2.46\pm0.59$} & {\boldmath $+0.18\pm0.043$} & {\boldmath $4.2$} \\
2227$-$088\tablenotemark{a} & PKS & $1.561$ & Q &  $848$ & $1.7$ & \nodata &  $< 0.12$ & \nodata \\
2230$+$114\tablenotemark{a} & CTA 102 & $1.037$ & Q &  $1380$ & $9.2$ & $-1.27\pm0.61$ & $-0.09\pm0.044$ & $2.1$ \\
2234$+$282\tablenotemark{a} & & $0.795$ & Q &  $820$ & $1.3$ & $-0.71\pm0.40$ & $< 0.14$ & \nodata \\
{\bf 2243$-$123} & {\bf PKS} & {\boldmath $0.630$} & {\bf Q} & {\boldmath $1705$} & {\boldmath $\sim0.2$}
                 & {\boldmath $-2.31\pm0.74$}  & {\boldmath $-0.14\pm0.043$} & {\boldmath $3.1$} \\
2251$+$158\tablenotemark{a} & 3C\,454.3 & $0.859$ & Q & $7638$ & $0.9$ & $+6.27\pm2.91$ & $+0.08\pm0.038$ & $2.2$ \\   
2345$-$167\tablenotemark{a} & PKS & $0.576$ & Q & $1457$ & $4.4$ & $+0.93\pm0.64$\tablenotemark{b} & $< 0.11$ & \nodata \\
\enddata
\tablecomments{Entries with CP detections at $\geq 3\sigma$ are in bold face. 
Object classes are from \citet{HB93}, BL = BL Lacerate Object (BL Lac) and  Q = Quasar.}
\tablenotetext{a}{Source used to make the smoothed $R/L$ gain corrections.}
\tablenotetext{b}{Similar strength noise elsewhere in image.}
\tablenotetext{c}{The CP peak is a bit south in the core and the image appears noisy. There also appears to be an approximately anti-symmetric +CP signal of 0.9 mJy/beam north of the core. }
\end{deluxetable}

\begin{deluxetable}{ccccccc}
\tablewidth{0pc}
\tablecolumns{7}
\tabletypesize{\scriptsize}
\tablenum{2}
\tablecaption{Uncertainties from Gain Transfer.\label{t:errors}}
\tablehead{$Object$ & $N_{scans}$ & $N_{cal}$ & $\sigma_{ran}$
& $\sigma_{avg}$ & $\sigma_{res}$ & $\sigma_{gains}$ \\
(B1950) &&& (\%) & (\%) & (\%) & (\%) }
\startdata
0215$+$015 & $15.5$ & $4.3$ & $0.059$ & $0.018$ & $0.031$ & $0.069$  \\
0219$-$164 & $16.1$ & $4.5$ & $0.051$ & $0.018$ & $0.031$ & $0.062$ \\
0219$+$428 & $18.0$ & $5.2$ & $0.034$ & $0.017$ & $0.029$ & $0.047$  \\
0336$-$019 & $13.8$ & $3.2$ & $0.038$ & $0.019$ & $0.036$ & $0.056$  \\
0403$-$132 & $13.1$ & $2.9$ & $0.039$ & $0.020$ & $0.038$ & $0.058$ \\
0420$-$014\tablenotemark{a} & $12.8$ & $2.8$ & $0.025$ & $0.020$ & $0.039$ & $0.050$ \\
0422$-$004 & $12.8$ & $2.8$ & $0.054$ & $0.020$ & $0.039$ & $0.069$ \\
0454$-$234 & $11.7$ & $2.7$ & $0.031$ & $0.021$ & $0.040$ & $0.055$ \\
0605$-$085\tablenotemark{a} & $11.9$ & $3.2$ & $0.023$ & $0.020$ & $0.037$ & $0.048$ \\
0607$-$157 & $11.3$ & $2.8$ & $0.022$ & $0.021$ & $0.039$ & $0.049$ \\
0743$-$006 & $12.9$ & $4.1$ & $0.027$ & $0.020$ & $0.032$ & $0.046$ \\
0846$+$513 & $15.1$ & $4.1$ & $0.086$ & $0.018$ & $0.032$ & $0.094$  \\
0851$+$202\tablenotemark{a} & $15.4$ & $4.3$ & $0.029$ & $0.018$ & $0.031$ & $0.046$ \\
0855$+$143 & $16.0$ & $4.8$ & $0.254$ & $0.018$ & $0.030$ & $0.256$  \\
0906$+$015 & $15.2$ & $4.6$ & $0.058$ & $0.018$ & $0.030$ & $0.068$ \\
0906$+$430 & $16.2$ & $4.7$ & $0.038$ & $0.018$ & $0.030$ & $0.052$ \\
1055$+$018\tablenotemark{a} & $19.3$ & $4.0$ & $0.027$ & $0.016$ & $0.033$ & $0.045$ \\
1150$+$497 & $18.2$ & $4.2$ & $0.045$ & $0.017$ & $0.032$ & $0.058$ \\
1156$+$295\tablenotemark{a} & $18.4$ & $3.6$ & $0.033$ & $0.016$ & $0.034$ & $0.050$ \\
1253$-$055\tablenotemark{a} & $18.2$ & $3.0$ & $0.022$ & $0.017$ & $0.037$ & $0.046$ \\
1334$-$127 & $16.8$ & $2.7$ & $0.025$ & $0.017$ & $0.040$ & $0.050$ \\
1404$+$286\tablenotemark{a} & $16.6$ & $3.1$ & $0.024$ & $0.017$ & $0.037$ & $0.047$ \\
1413$+$135 & $13.3$ & $2.4$ & $0.034$ & $0.019$ & $0.042$ & $0.057$ \\
1502$+$106 & $13.2$ & $2.5$ & $0.030$ & $0.019$ & $0.041$ & $0.055$ \\
1504$-$166 & $13.7$ & $2.4$ & $0.030$ & $0.019$ & $0.042$ & $0.055$ \\
1510$-$089 & $13.1$ & $2.5$ & $0.062$ & $0.020$ & $0.041$ & $0.077$ \\
1546$+$027 & $12.9$ & $2.6$ & $0.022$ & $0.020$ & $0.040$ & $0.050$ \\
1656$+$053 & $12.5$ & $3.6$ & $0.036$ & $0.020$ & $0.034$ & $0.053$ \\
1842$+$681 & $16.6$ & $4.7$ & $0.043$ & $0.017$ & $0.030$ & $0.055$ \\
1921$-$293 & $15.1$ & $3.4$ & $0.023$ & $0.018$ & $0.035$ & $0.046$ \\
1958$-$179\tablenotemark{a} & $18.7$ & $4.1$ & $0.026$ & $0.016$ & $0.032$ & $0.045$ \\
2032$+$107 & $19.7$ & $4.6$ & $0.032$ & $0.016$ & $0.030$ & $0.047$ \\
2201$+$171 & $22.6$ & $5.5$ & $0.045$ & $0.015$ & $0.028$ & $0.055$ \\
2223$-$052 & $23.5$ & $5.7$ & $0.027$ & $0.015$ & $0.027$ & $0.041$ \\
2227$-$088\tablenotemark{a} & $23.6$ & $5.5$ & $0.034$ & $0.015$ & $0.028$ & $0.046$ \\
2230$+$114\tablenotemark{a} & $22.9$ & $5.4$ & $0.027$ & $0.015$ & $0.028$ & $0.042$ \\
2234$+$282\tablenotemark{a} & $23.7$ & $5.4$ & $0.034$ & $0.015$ & $0.028$ & $0.046$ \\
2243$-$123 & $23.9$ & $5.6$ & $0.028$ & $0.014$ & $0.027$ & $0.042$ \\
2251$+$158\tablenotemark{a} & $23.5$ & $5.5$ & $0.021$ & $0.015$ & $0.028$ & $0.038$ \\
2345$-$167\tablenotemark{a} & $22.7$ & $5.3$ & $0.027$ & $0.015$ & $0.028$ & $0.042$ \\
\enddata
\tablecomments{$N_{scan}$ is the mean number of scans which contribute to the
antenna gain averages applied to a source. $N_{cal}$ is the weighted
number of calibrator sources which determine the antenna gains averages
applied to a source, see description in text.}
\tablenotetext{a}{Source used to make the smoothed $R/L$ gain corrections.}
\end{deluxetable}


\begin{thebibliography}{dummy}

\bibitem[Attridge (1999)]{A99}
Attridge, J. M. 1999
Ph.D. Thesis, Brandeis Univ.






\bibitem[Blandford \& Payne(1982)]{BP82}
Blandford, R. D., \& Payne, D. G. 1982
\mnras, 199, 883


\bibitem[Bower, Falcke \& Backer(1999)]{BFB99}
Bower, G. C., Falcke, H., \& Backer, D. C. 1999
\apj, 523, L29

\bibitem[Bridle \& Greisen (1994)]{BG94}
Bridle, A. H., \& Greisen, E. W. 1994 AIPS Memo 87




\bibitem[Cotton(1993)]{C93} 
Cotton, W. D. 1993 \aj, 106, 1241




\bibitem[Fender et al.(2000)]{F00}
Fender, R., Rayner, D., Norris, R., Sault, R. J., \& Pooley, G. 2000
\apj, 530, L29











\bibitem[Greisen (1988)]{G88}
Greisen, E. W. 1988 AIPS Memo 61

\bibitem[Hewitt \& Burbidge(1993)]{HB93}
Hewitt, A., \& Burbidge, G. 1993 \apjs, 87, 451


\bibitem[Homan et al.(2001)]{H01}
Homan, D. C., Ojha, R., Wardle, J. F. C., Roberts, D. H.,
Aller, M. F., Aller, H. D., \& Hughes, P. A. 2001 \apj, in press, 
astro-ph/0009301

\bibitem[Homan \& Wardle(1999)]{HW99} 
Homan, D. C., \& Wardle, J. F. C. 1999 \aj, 118, 1942



\bibitem[Hodge(1982)]{H82}
Hodge, P. E. 1982 \apj, 263, 595

\bibitem[Hodge \& Aller(1977)]{HA77}
Hodge, P. E., \& Aller, H. D. 1977 \apj, 211, 669




\bibitem[Jones(1988)]{J88}
Jones, T. W. 1988 \apj, 332, 678

\bibitem[Jones \& O'Dell(1977)]{JOD77}
Jones, T. W., \& O'Dell, S. L. 1977 \apj, 214, 522



\bibitem[Komesaroff et al.(1984)]{K84}
Komesaroff, M. M., Roberts, J. A., Milne, D. K., 
Rayner, P. T., \& Cooke, D. J. 1984 \mnras, 208, 409 




\bibitem[Lovelace \& Romanova(1995)]{LR95}
Lovelace, R. V. E., \& Romanova, M. M. 1995 in 
ASP Conf. Ser. 100, Energy Transport in Radio Galaxies and Quasars,
ed. Hardee, P. E., Bridle, A. H., \& Zensus, J. A., 25

\bibitem[Macquart et al.(2000)]{M00}
Macquart, J.-P., Kedziora-Chudczer, L., Rayner, D. P., Jauncey, D. L. 2000
\apj, 538, 623

\bibitem[Macquart \& Melrose(2000)]{MM00}
Macquart, J.-P., \& Melrose, D. B. 2000
\apj, 545, 798





\bibitem[Napier(1995)]{N95} 
Napier, P.J. 1995 in ASP Conf. Ser. 82, Very Long Baseline 
Interferometry with the VLBA, ed. Zensus, J. A., Diamond, 
P. J., \& Napier, P. J., 57 




\bibitem[Rayner, Norris \& Sault(2000)]{RNS00}
Rayner, D. P., Norris, R. P., \& Sault, R. J. 2000 \mnras, 319, 484


\bibitem[Roberts, Wardle \& Brown(1994)]{RWB94} 
Roberts, D. H., Wardle, J. F. C., \& Brown, L. F. 1994, \apj, 427, 718



\bibitem[Sault \& Macquart(1999)]{SM99}
Sault, R. J., \& Macquart, J.-P. 1999
\apj, 526, L85





\bibitem[Taylor(1998)]{T98}
Taylor, G. B. 1998 \apj, 506, 637

\bibitem[Taylor(2000)]{T00}
Taylor, G. B. 2000 \apj, 533, 95






\bibitem[Wardle et al.(1998)]{WHOR98} 
Wardle, J. F. C., Homan, D. C., Ojha, R., \& Roberts, D. H. 1998 
\nat, 395, 457

\bibitem[Wardle \& Homan(2001)]{WH01}
Wardle, J. F. C., \& Homan, D. C. 2001
in ASP Conf. Ser. (in press), Particles and Fields in Radio
Galaxies, ed. Blundell, K. M., \& Laing, R. A., astro-ph/0011515



\bibitem[Weiler \& de Pater(1983)]{WDP83}
Weiler, K. W., \& de Pater, I. 1983 \apjs, 52, 293




\end{thebibliography}
\end{document}